\documentclass[11pt]{article}

\usepackage{color}

\usepackage{amsfonts,amssymb,amsthm}
\usepackage{amsmath,amssymb}
\usepackage{mathrsfs}
\usepackage{enumerate}
\usepackage{srcltx}
\usepackage[ansinew]{inputenc}
\usepackage{mathtools}
    \mathtoolsset{showonlyrefs}
    \mathtoolsset{centercolon}
\usepackage{graphicx}

\DeclareMathAlphabet{\mathbf}{OT1}{cmr}{bx}{it}
\DeclareMathAlphabet{\mathssb}{OT1}{cmss}{bx}{n}
\DeclareMathAlphabet{\mathssn}{OT1}{cmss}{m}{n}
\DeclareMathAlphabet{\mathub}{OT1}{cmr}{b}{n}

\newcommand\jump[1]{[\![#1]\!]}

\newcommand {\ab} {\mathbf{a}}
\newcommand {\bb} {\mathbf{b}}
\newcommand {\cb} {\mathbf{c}}
\newcommand {\qb} {\mathbf{q}}
\newcommand {\hb} {\mathbf{h}}
\newcommand {\ub} {\mathbf{u}}
\newcommand {\vb} {\mathbf{v}}
\newcommand {\fb} {\mathbf{f}}
\newcommand {\eb} {\mathbf{e}}
\newcommand {\nb} {\mathbf{n}}
\newcommand {\wb} {\mathbf{w}}
\newcommand {\Fb} {\mathbf{F}}
\newcommand {\Qb} {\mathbf{Q}}
\newcommand {\Wb} {\mathbf{W}}
\newcommand {\Sc}  {\mathcal{S}}
\newcommand {\Vc}  {\mathcal{V}}
\newcommand {\Ib} {\mathbf{I}}
\newcommand {\sym} {\textrm{sym}}
\renewcommand {\div} {\textrm{div}}

\newcommand\remarkname{Remark} %
\newcounter {remarkn}[section]%
\renewcommand \theremarkn {\arabic{section}.\arabic{remarkn}}%
\newcommand\remarksname{Remark} %
\newcounter {remarksn}[section]%
\renewcommand \theremarksn {\thesection.\arabic{remarksn}}%

\begin{document}

\begin{center}

\end{center}

\newcommand\email[1]{\texttt{#1}}
\newcommand\at{:}

\begin{center}
 {\bf \Large
On Energy and Entropy Influxes in the Green--Naghdi Type III Theory of Heat Conduction}
\end{center}
\medskip

\begin{center}
{\large  Swantje Bargmann$^\star$ \quad
        Antonino Favata$^\diamond$ \quad
        Paolo Podio-Guidugli$^\diamond$
}\end{center}

\begin{center}
 \noindent $^\star$ Institute of Continuum Mechanics and Materials Mechanics, Hamburg University of Technology\footnote{Eißendorfer Str. 42
 21073 Hamburg,
 Germany.
  \\
 {\null} \quad \ Email:
 \begin{minipage}[t]{30em}
  \email{swantje.bargmann@tu-harburg.de} (S. Bargmann)
 \end{minipage}} \& Institute of Materials Research, Helmholtz-Zentrum\footnote{Max-Planck-Straße 1,  21502 Geesthacht, Germany.} 
 
  \noindent $^\diamond$Dipartimento di Ingegneria Civile e Ingegneria Informatica, Universit\`a di Roma Tor Vergata\footnote{Via Politecnico 1, 00133 Rome, Italy. \\
{\null} \quad \ Email:
\begin{minipage}[t]{30em}
\email{favata@ing.uniroma2.it} (A. Favata)\\
\email{ppg@uniroma2.it} (P. Podio-Guidugli)
\end{minipage}}
\small
\end{center}
\medskip

\begin{abstract}
\noindent {\footnotesize The energy-influx$\,/\,$entropy-influx relation in the Green--Naghdi Type III theory of heat conduction is examined within a  thermodynamical framework \emph{\`a la} M\"uller--Liu, where that relation is not specified a priori irrespectively of the constitutive class under attention. It is shown that the classical assumption, i.e.,  that the entropy influx and the energy influx  are proportional via the absolute temperature, holds true if heat conduction is, in a sense that is made precise, isotropic. In addition, it is proven that  the standard assumption does not hold in case of transversely isotropic conduction.

\medskip

\noindent\textbf{Keywords:}\ {energy influx, entropy influx, M\"uller--Liu entropy principle, coldness, thermal displacement}}
\end{abstract}

\section{Introduction}
A key  assumption in the standard continuum theory of heat propagation is that, for $\,\qb,\hb$ and $\,r,s$ the influx vectors and the external sources of, respectively, energy and entropy, the energy inflow $\,(\qb,r)$
 is proportional to the entropy inflow $\,(\hb,s)$, the proportionality factor being the absolute temperature $\vartheta$:
\begin{equation}\label{prop}
(\qb,r)=\vartheta(\hb,s),\quad\vartheta>0.
\end{equation}
Whether or not this assumption is generally tenable has been the subject of considerable debate (see e.g. the papers by M\"uller \cite{Mu} and Liu \cite{Liu,Liu1}). The issue can be taken up in the framework  of one or another continuum theory of heat conduction: Podio-Guidugli \cite{PPG} uses the framework of the standard theory, which leads to a parabolic heat equation, Bargmann and Steinmann \cite{BS} that of the so-called Type III theory of Green and Naghdi \cite{GN}, which allows for propagation of heat waves with finite speed (we summarize these two theories in Section 2). In this paper, we use the latter framework, in a manner that differs from Bargmann and Steinmann's in a number of points. As in \cite{PPG} for the standard theory, we prove: in Section 3,  that for thermodynamic consistency \eqref{prop} must hold true when heat conduction is, in a sense that we make precise, \emph{isotropic}; in Section 4,  that \eqref{prop} may be violated in the \emph{transversely isotropic} case. In our proofs, we exploit certain  mathematical tools that are described in Appendices A and B.

\section{Heat conduction theories }
\noindent Thermal and deformational phenomena can be considered coupled, as is done, e.g., in thermoelasticity. For our present purposes, it is sufficient to consider the simplest instance, namely, heat propagation in a rigid conductor. Consequently, we need not distinguish between reference and current configurations; we also leave all spatial dependences tacit.

The standard continuum theory of heat conduction is based on the following two laws:

\noindent -- (\emph{energy balance})
\begin{equation}\label{eneq}
 \rho\,\dot\varepsilon=-\div\qb+ \rho\, r,
\end{equation}
where $ \rho\,$ denotes the volumetric mass density, $\varepsilon$ the internal energy per unit volume, and $\dot\varepsilon$ its time rate;

\noindent -- (\emph{entropy imbalance})
\begin{equation}\label{enineq}
 \rho\,\dot\eta\geq-\div\hb+ \rho\, s,
\end{equation}
with $\eta$ the entropy. It is customarily assumed that \eqref{prop} holds.
With this and the notion of \emph{Helmholtz free-energy} per unit mass:
\[
\psi=\varepsilon-\vartheta\eta,
\]
the following \emph{free-energy growth inequality} is arrived at:
\begin{equation}\label{fein}
\dot\psi\leq -\eta\dot\vartheta-\vartheta^{-1}\,\qb\cdot\nabla\vartheta\,.
\end{equation}
When a set of state variables is chosen -- say, $(\vartheta,\nabla\vartheta,\dot\vartheta)$ --
this inequality is used to derive restrictions on the choice of the state functions delivering $\psi$, $\eta$, and $\qb$, in the manner devised by Coleman and Noll in their classic paper \cite{CN}.

Green \& Naghdi modified the classic path, in that they took the following entropy law as their point of departure:

\noindent -- (\emph{entropy balance})
\begin{equation}\label{CD}
 \rho\,\xi= \rho\,\dot\eta+\div\hb- \rho\, s, 
\end{equation}
where $\xi$ is the \emph{internal entropy production}. As to internal energy, they accepted the balance equation \eqref{eneq}, implicitly excluding any internal production of energy; as to entropy and energy inflows, the classic proportionality expressed by \eqref{prop}. Consequently, they arrived at the following version of \eqref{CD} in terms of Helmholtz free energy:
\begin{equation}\label{CDG}
\dot\psi+\eta\dot\vartheta+\hb\cdot\nabla\vartheta+\xi=0.
\end{equation}
 Moreover, in their so-called Type III theory, they assumed that  the constitutive mappings delivering $\psi,\eta$, and $\hb$, depend on the following list of state variables:
\begin{equation}\label{statvar}
\Sc:= \{\alpha, \dot\alpha, \nabla\alpha, \nabla\dot\alpha\},
\end{equation}
where the \textit{thermal displacement} $\alpha$, a field whose precise interpretation in statistical mechanics is still wanted, is somehow indirectly characterized by postulating that its time derivative equals an \emph{empirical temperature} $T$:
\[
\dot\alpha(t):=T(t).\footnote{Here is a quotation from \cite{GNRoy}: ``The temperature $T$ (on the macroscopic
scale) is generally regarded as representing (on thermolecular scale) some `mean' velocity magnitude or `mean'
(kinetic energy)$^{1/2}$. With this in mind, we introduce a scalar $\alpha=\alpha(X,t)$ through an integral of the form
\[
\qquad\qquad\qquad\qquad\qquad\qquad\quad\alpha=\int_{t_0}^t T(X,\tau)\,d\tau +\alpha_0,\;\qquad\qquad\qquad\qquad\qquad\qquad(7.3)
\]
where $t_0$ denotes some reference time and the constant $\alpha_0$ is the initial value of $\alpha$ at time $t_0$. In view of the above interpretation associated with $T$ and the physical dimension of the quantity defined by (7.3), the variable
$\alpha$ may justifiably be called thermal displacement magnitude or simply thermal displacement. Alternatively,
we may regard the scalar $\alpha$  (on the macroscopic scale) as representing a `mean' displacement magnitude on
the molecular scale and then $T = \dot\alpha$.''}
\]

As anticipated in the Introduction, M\"uller and Liu questioned the general applicability of assumption \eqref{prop} and proposed a thermodynamic format within which to test whether or not it fits a given constitutive class. This format amounts to considering the energy balance \eqref{eneq} as an internal constraint to be appended to the entropy imbalance \eqref{enineq} after multiplication by a so-called `Lagrange multiplier' $\lambda$, a positive scalar to be constitutively specified:
\begin{equation}\label{entrimb}
 \rho\,\dot\eta+\div\hb- \rho\, s-\lambda\big(  \rho\,\dot\varepsilon+\div\qb- \rho\, r\big)\geq0,\quad \lambda>0.
\end{equation}
When the constitutive mappings involved in \eqref{entrimb} depend on the Green-Naghdi list \eqref{statvar} of state variables, one can try and use a Coleman-Noll type procedure to see whether the multiplier $\lambda$ may be shown to be, if not equal, at least proportional to \emph{coldness} \cite{Mu}, that is, the inverse $\vartheta^{-1}$ of the absolute temperature. 

Granted an appropriate generalization of the format we just introduced, the same issue has been considered by Bargmann and Steinmann in \cite{BS} for thermoelastic materials. We here reach more complete conclusions  than theirs under less stringent assumptions and by a different train of reasoning.

\section{Isotropic conduction implies proportionality of energy and entropy influxes}
On choosing the state variables as in \eqref{statvar}, inequality \eqref{entrimb} reads:
\begin{equation}\label{diseq}
\begin{aligned}
& \rho\,\big(\partial_\alpha\eta-\lambda\partial_\alpha\varepsilon \big)\dot\alpha+ \rho\,\big(\partial_{\dot\alpha}\eta-\lambda\partial_{\dot\alpha}\varepsilon \big)\ddot\alpha+ \rho\,\big(\partial_{\nabla\alpha}\eta-\lambda\partial_{\nabla\alpha}\varepsilon \big)\cdot\nabla\dot\alpha+\\
& \rho\,\big(\partial_{\nabla\dot\alpha}\eta-\lambda\partial_{\nabla\dot\alpha}\varepsilon \big)\nabla\ddot\alpha+\div\hb-\lambda\div\qb\geq0,
\end{aligned}
\end{equation}
where
\begin{equation}\label{exc}
\begin{aligned}
\div\hb-\lambda\div\qb&=\big(\partial_\alpha\hb-\lambda\,\partial_\alpha\qb \big)\cdot\nabla\alpha+\big(\partial_{\dot\alpha}\hb-\lambda\,\partial_{\dot\alpha}\qb \big)\cdot\nabla\dot\alpha+\\
&+\big(\partial_{\nabla\alpha}\hb-\lambda\,\partial_{\nabla\alpha}\qb \big)\cdot\nabla^2\alpha+\big(\partial_{\nabla\dot\alpha}\hb-\lambda\,\partial_{\nabla\dot\alpha}\qb \big)\cdot\nabla^2\dot\alpha.
\end{aligned}
\end{equation}

 We apply the Coleman-Noll procedure \cite{CN} to the inequality that obtains by combining \eqref{diseq} and \eqref{exc}, that is, we require that it be  satisfied whatever the local continuation $(\ddot\alpha, \nabla\ddot\alpha, \nabla^2\alpha, \nabla^2\dot\alpha)$
of each admissible process. This requirement is satisfied if and only if 
\begin{equation}\label{cond}
\begin{aligned}
&\sym\big(\partial_{\nabla\alpha}\hb-\lambda\,\partial_{\nabla\alpha}\qb \big)=\mathbf{0},\\
&\sym\big(\partial_{\nabla\dot\alpha}\hb-\lambda\,\partial_{\nabla\dot\alpha}\qb \big)=\mathbf{0},\\
&\partial_{\dot\alpha}\eta-\lambda\,\partial_{\dot\alpha}\varepsilon=0,\\
&\partial_{\nabla\dot\alpha}\eta-\lambda\,\partial_{\nabla\dot\alpha}\varepsilon=0,
\end{aligned}
\end{equation}
and, moreover, inequality \eqref{entrimb} reduces to 
\begin{equation}
\begin{aligned}
& \rho\,\big(\partial_\alpha\eta-\lambda\,\partial_\alpha\varepsilon \big)\dot\alpha+\big( \rho\,\big(\partial_{\nabla\alpha}\eta-\lambda\,\partial_{\nabla\alpha}\varepsilon \big)+\partial_{\dot\alpha}\hb-\lambda\,\partial_{\dot\alpha}\qb\big)\cdot\nabla\dot\alpha+\big(\partial_\alpha\hb-\lambda\,\partial_{\alpha}\qb \big)\cdot\nabla\alpha\geq0.
\end{aligned}
\end{equation}

Conditions $\eqref{cond}_{1,2}$ are expedient to prove our main result: 

\begin{quote}\emph{For each chosen value of the  independent variables $\alpha,\dot\alpha$, let the energy and entropy influxes and the Lagrangian multiplier be delivered  by \emph{isotropic} constitutive mappings $\widehat\qb(\alpha,\dot\alpha,\cdot,\cdot)$, $\widehat\hb(\alpha,\dot\alpha,\cdot,\cdot)$, $\widehat\lambda(\alpha,\dot\alpha,\cdot,\cdot)$, the first two vector-valued, the third delivering positive scalars. Then,  the energy influx is proportional to the entropy influx via the Lagrangian multiplier, which depends neither on $\nabla\alpha$ nor on $\nabla\dot\alpha$:}
\begin{equation}\label{qproph}
\hb=\lambda\qb,\quad \lambda=\widehat\lambda(\alpha,\dot\alpha)>0.
\end{equation}
\end{quote}
Our proof is achieved as follows. For simplicity, we leave the dependence on $(\alpha,\dot\alpha)$ tacit, and write $\ub$ for the third state variable in the list \eqref{statvar}, and $\vb$ for the fourth. On making use of  the representation \eqref{repr} obtained in Appendix A, we set:
\begin{equation}\label{repre}
\begin{aligned}
&\widehat{\hb}(\ub,\vb)=h_1\ub+h_2\vb,\quad h_j=\widehat h_j(|\ub|,|\vb|,\ub\cdot\vb)\;(j=1,2),\\
&\widehat{\qb}(\ub,\vb)=q_1\ub+q_2\vb,\quad q_j=\widehat q_j(|\ub|,|\vb|,\ub\cdot\vb)\;(j=1,2),
\end{aligned}
\end{equation}
whence
\begin{equation}\label{prelim}
\hb-\lambda\qb=(h_1-\lambda q_1)\ub+(h_2-\lambda q_2)\vb.
\end{equation}
With this, condition $\eqref{cond}_1$ reads:
\begin{equation}\label{cond1}
\begin{aligned}
&\big(h_1-\lambda q_1 \big)\Ib+\big(\partial_1 h_1-\lambda\partial_1q_1 \big)|\ub|^{-1}\,\ub\otimes\ub+\big(\partial_3 h_2-\lambda\partial_3q_2 \big)\,\vb\otimes\vb+\\
&\Big(\partial_3h_1+\partial_1h_2|\ub|^{-1}-\lambda\big(\partial_3q_1+\partial_1q_2|\ub|^{-1}\big)  \Big)\,\sym\big(\ub\otimes\vb+\vb\otimes\ub\big)=\mathbf{0},
\end{aligned}
\end{equation}
and has to be satisfied for all $\ub$ and $\vb$ (here $\partial_i\;( i=1,2,3)$ denotes differentiation of a function with respect to the $i$-th of its arguments). An application of the first of the two algebraic lemmas proved in Appendix B permits us to conclude that:
\begin{equation}\label{ccond}
\begin{aligned}
&h_1-\lambda q_1=0,\\
&\partial_1\big(h_1-\lambda q_1\big)+\big(\partial_1\lambda\big) q_1=0,\\
&\partial_3h_2-\lambda\,\partial_3 q_2=0,\\
&\partial_3 h_1-\lambda\,\partial_3 q_1=0,\\&\partial_1 h_2-\lambda\, \partial_1 q_2=0.
\end{aligned}
\end{equation}
Quite analogously, condition $\eqref{cond}_2$ yields:
\begin{equation}\label{ccond1}
\begin{aligned}
&h_2-\lambda q_2=0,\\
&\partial_2\big(h_2-\lambda q_2\big)+\big(\partial_2\lambda\big) q_2=0,\\
&\partial_3h_1-\lambda\,\partial_3 q_1=0,\\
&\partial_3 h_2-\lambda\,\partial_3 q_2=0,\\&\partial_2 h_2-\lambda\, \partial_2q_2=0.
\end{aligned}
\end{equation}
Now, \eqref{ccond} and \eqref{ccond1} imply that the Lagrange multiplier $\lambda$ cannot depend on $\ub, \vb$, that is to say, that
\begin{equation}
\lambda=\widehat{\lambda}(\alpha,\dot\alpha);
\end{equation}
with this, $\eqref{ccond}_1$, and $\eqref{ccond1}_1$, \eqref{prelim} becomes
\begin{equation}\label{propo}
\widehat\hb(\alpha,\dot\alpha,\nabla\alpha,\nabla\dot\alpha)=\widehat\lambda(\alpha,\dot\alpha)\,\widehat\qb(\alpha,\dot\alpha,\nabla\alpha,\nabla\dot\alpha),
\end{equation}
thus, $\eqref{qproph}_1$ is established.

A direct consequence of \eqref{propo} is that, when two material bodies are in \emph{ideal thermal contact}, that is, by definition, when neither the energy nor the entropy flux suffers a jump at a point of a common interface oriented by the normal field $\nb$:
\begin{equation}
\jump{\qb\cdot\nb}=\jump{\hb\cdot\nb}=0,
\end{equation}
then the Lagrange multiplier is also continuous at the interface:
\begin{equation}
\jump{\lambda}=0.
\end{equation}

At this point, in the words of M\"uller, ``\dots there is a price to pay for not having introduced the temperature so far in rational thermodynamics with Lagrange multipliers" (\cite{ITD}, p. 168). The \emph{desideratum} is a proof that 
\begin{equation}\label{desi}
\lambda \propto \vartheta^{-1}, 
\end{equation}
with $\vartheta^{-1}$ the \emph{coldness}; hence, in particular, that $\widehat\lambda$ is a universal function. This can be achieved, as M\"uller himself proposed in \cite{Mu}, by considering a situation when a material body of whatsoever constitutive nature is put in ideal thermal contact with an ideal gas, for which the kinetic theory permits to show that \eqref{desi} indeed holds, provided $\vartheta$ is taken proportional to the empirical temperature $T$. We leave it to the reader to decide whether or not such an argument is convincing. For us, in that it relies on importing a result from a theory tacitly regarded as more foundational in nature, it is for sure suggestive  of assuming \eqref{desi} right away, without any need to imply, let alone accept, a vassalage between theories.

\section{A counterexample to proportionality: transversely isotropic conduction}
In this Section, we show that assumption \eqref{prop} is not tenable in general. Within the framework of a standard theory of heat conduction modified by the introduction of a Lagrangian multiplier,  this result has been announced  in \cite{Liu2} by Liu, who proposed to consider the case of transversely isotropic materials to exhibit a counterexample to proportionality of energy and entropy influxes; the proof he offered is faulty, although amendable as indicated  in \cite{PPG}. We here exploit the same counterexample within the Green-Naghdi theory of Type III.

For $\eb\in\Vc$ any chosen unit vector, consider the class of vector-valued mappings over $\Vc\times\Vc$ of the following form:
\begin{equation}\label{tisomap}
\widetilde{\fb}(\ub,\vb;\eb)=\varphi_1\,\ub+\varphi_2\,\vb+\varphi_3\,\eb,\quad \varphi_i=\widetilde\varphi_i(|\ub|, |\vb|, \ub\cdot\vb, \ub\cdot\eb,\vb\cdot\eb)\;(i=1,2,3)
\end{equation}
(once again, we have left the dependence on $(\alpha,\dot\alpha)$ tacit). Clearly, each mapping $\widetilde\fb$ in this class is \emph{transversely isotropic} with respect to the axis $\eb$, in the sense that it satisfies
\begin{equation}\label{tiso}
\Qb\widetilde{\fb}(\ub,\vb)=\widetilde{\fb}(\Qb\ub,\Qb\vb), \quad\forall\,\ub,\vb\in \Vc, \;\;\forall\,\Qb\in{\rm Orth}(\eb),
\end{equation}
where ${\rm Orth}(\eb)$ denotes the continuous group of all rotations about the span of $\eb$.

On using a representation if type \eqref{tisomap} for both  $\widetilde{\hb}$ and $\widetilde{\qb}$:
\begin{equation}\label{repret}
\begin{aligned}
\widetilde{\hb}(\ub,\vb)&=h_1\,\ub+h_2\,\vb+h_3\,\eb,\quad h_i=\widetilde h_i(|\ub|, |\vb|, \ub\cdot\vb, \ub\cdot\eb,\vb\cdot\eb)\;(i=1,2,3),\\
\widetilde{\qb}(\ub,\vb)&=q_1\,\ub+q_2\,\vb+q_3\,\eb,\quad h_i=\widetilde q_i(|\ub|, |\vb|, \ub\cdot\vb, \ub\cdot\eb,\vb\cdot\eb)\;(i=1,2,3),\\
\end{aligned}
\end{equation}
whence
\begin{equation}
\hb-\lambda\qb=(h_1-\lambda q_1)\ub+(h_2-\lambda q_2)\vb+(h_3-\lambda q_3)\eb.
\end{equation}
Now, condition $\eqref{cond}_1$ becomes:
\begin{equation}\label{cond1t}
\begin{aligned}
&\big(h_1-\lambda q_1 \big)\Ib+\big(\partial_1 h_1-\lambda\partial_1q_1 \big)|\ub|^{-1}\,\ub\otimes\ub+\big(\partial_3 h_2-\lambda\partial_3q_2 \big)\,\vb\otimes\vb+\\
&\big(\partial_4h_3-\lambda\partial_4 q_3 \big)\,\eb\otimes\eb+\\
&\Big(\partial_3h_1+\partial_1h_2|\ub|^{-1}-\lambda\big(\partial_3q_1+\partial_1q_2|\ub|^{-1}\big)  \Big)\,\sym\big(\ub\otimes\vb+\vb\otimes\ub\big)+\\
&\Big(\partial_4h_1+\partial_1h_3|\ub|^{-1}-\lambda\big( \partial_4q_1+\partial_1q_3|\ub|^{-1}\big) \Big)\,\sym\big(\ub\otimes\eb+\eb\otimes\ub\big)+\\
&\Big(\partial_4h_2+\partial_3h_3-\lambda\big( \partial_4q_2+\partial_3q_3\big) \Big)\,\sym\big(\vb\otimes\eb+\eb\otimes\vb\big)=\mathbf{0},
\end{aligned}
\end{equation}
for all $\ub$ and $\vb$; condition $\eqref{cond}_2$ yields a completely analogous identity.
By applying Lemma 2 in Appendix B to each of these two identities, we recover relations \eqref{ccond} and \eqref{ccond1}, as well as the following list of additional relations:
\begin{equation}\label{ccondt}
\begin{aligned}
&\partial_4 h_3-\lambda\,\partial_4 q_3=0,\\
&\partial_4 h_1-\lambda\,\partial_4 q_1=0,\\
&\partial_1 h_3-\lambda\, \partial_1 q_3=0,\\
&\partial_4 h_2-\lambda\,\partial_4 q_2+\partial_3 h_3-\lambda\,\partial_3 q_3=0,
\end{aligned}
\end{equation}
and 
\begin{equation}\label{ccond1t}
\begin{aligned}
&\partial_5 h_3-\lambda\, \partial_5q_3=0,\\
&\partial_5 h_2-\lambda\,\partial_5 q_2=0,\\
&\partial_2 h_3-\lambda\,\partial_2 q_3=0,\\
&\partial_5 h_1-\lambda\,\partial_5 q_1+\partial_3 h_3-\lambda\,\partial_3 q_3=0,
\end{aligned}
\end{equation}
whence
\begin{equation}
h_1-\lambda q_1=0,\quad h_2-\lambda q_2=0, \quad h_3-\lambda q_3=f \quad \textrm{with}\quad f=\widetilde f(\alpha,\dot\alpha),\;\,\lambda=\widetilde\lambda(\alpha,\dot\alpha).
\end{equation}
In conclusion, \emph{an influx discrepancy in the direction of the transverse isotropy remains}:
\begin{equation}
\hb-\lambda\qb=(h_3-\lambda q_3)\eb=f\,\eb,
\end{equation}
where
\[
\begin{aligned}
\widetilde h_3(\alpha,\dot\alpha,|\nabla\alpha|,|\nabla\dot\alpha|,&\nabla\alpha\cdot\nabla\dot\alpha,\partial_\eb\alpha,\partial_\eb\dot\alpha)=\\&\widetilde\lambda(\alpha,\dot\alpha)\,\widetilde q_3(\alpha,\dot\alpha,|\nabla\alpha|,|\nabla\dot\alpha|,\nabla\alpha\cdot\nabla\dot\alpha,\partial_\eb\alpha,\partial_\eb\dot\alpha)+\widetilde f(\alpha,\dot\alpha).
\end{aligned}
\]

\section*{Appendix A. Representation of an isotropic vector function of two vector arguments}\label{lemma}

Here we prove a representation formula for isotropic vector-valued mappings of two vector arguments. As a premiss, we recall that a
scalar-valued function $f$ of two vectorial arguments is \textit{isotropic}, i.e., such that
$$
f(\ub,\vb)=f(\Qb\ub,\Qb\vb), \quad \forall \ub,\vb \in \Vc, \;\;\forall\Qb\in{\rm Orth},
$$
where $\Vc$ is a suitable vector space and Orth the full orthogonal group, if and only if
\begin{equation}\label{isosca}
f(\ub,\vb)=\varphi(|\ub|, |\vb|, \ub\cdot\vb).
\end{equation}

\paragraph{Lemma 1}\label{lemmaa}Let $\widehat{\fb}$ be an \emph{isotropic} mapping  of $\Vc$ into itself, i.e., such that
\begin{equation}\label{iso}
\Qb\widehat{\fb}(\ub,\vb)=\widehat{\fb}(\Qb\ub,\Qb\vb), \quad\forall\ub,\vb\in \Vc, \;\;\forall\Qb\in{\rm Orth}.
\end{equation}
Then, $\widehat{\fb}$ has the following representation:
\begin{equation}\label{repr}
\widehat{\fb}(\ub,\vb)=\varphi_1(|\ub|, |\vb|, \ub\cdot\vb)\,\ub+\varphi_2(|\ub|, |\vb|, \ub\cdot\vb)\,\vb.
\end{equation}
\noindent\textit{Proof} Write \eqref{iso} for $\Qb$ a point of a smooth curve through $\Ib$ on the manifold Orth:
\begin{equation}
t\mapsto\Qb(t), \qquad \Qb(t_0)=\Ib, \qquad \dot\Qb(t_0)=:\Wb\in{\rm Skw},
\end{equation}
and differentiate at $t=t_0$, so as to get:
\begin{equation}
\begin{aligned}
\Wb\widehat{\fb}(\ub,\vb)=&\widehat{\Fb}_1(\ub,\vb)\Wb\ub+\widehat{\Fb}_2(\ub,\vb)\Wb\vb,\\
&\widehat{\Fb}_1(\ub,\vb):=\partial_1\widehat{\fb}(\ub,\vb), \;\widehat{\Fb}_2(\ub,\vb):=\partial_2\widehat{\fb}(\ub,\vb), \quad\forall\ub,\vb\in\Vc,\; \forall\Wb\in{\rm Skw}.
\end{aligned}
\end{equation}
This condition is equivalent to
\begin{equation}
\Wb\cdot\big(\ab\otimes\widehat{\fb}(\ub,\vb)-\widehat{\Fb}_1^T(\ub,\vb)\ab\otimes\ub-\widehat{\fb}(\ub,\vb)-\widehat{\Fb}_2^T(\ub,\vb)\ab\otimes\vb \big), \quad\forall\ub,\vb,\ab\in\Vc,\; \forall\Wb\in{\rm Skw},
\end{equation}
and then
\begin{equation}
\ab\times\widehat{\fb}(\ub,\vb)-\widehat{\Fb}_1^T(\ub,\vb)\ab\times\ub-\widehat{\Fb}_2^T(\ub,\vb)\ab\times\vb=\mathbf{0},  \quad\forall\ub,\vb,\ab\in\Vc.
\end{equation}
Now, take the vector product with $\wb:=\ub\times\vb$ and use the identity $(\ab\times\bb)\times\cb=(\ab\cdot\cb)\bb-(\bb\cdot\cb)\ab$ so as to obtain:
\begin{equation}
(\ab\cdot\wb)\widehat{\fb}(\ub,\vb)-\big( \widehat{\fb}(\ub,\vb)\cdot\wb\big)\ab=\big(\ab\cdot\widehat{\Fb}_1(\ub,\vb)\wb\big)\ub+\big(\ab\cdot\widehat{\Fb}_2(\ub,\vb)\wb\big)\vb \quad \quad\forall\ub,\vb,\ab\in\Vc. 
\end{equation}
Since 
\begin{equation}
\begin{aligned}
&\big(\ab\cdot\widehat{\Fb}_1(\ub,\vb)\wb\big)\ub=\big(\ub\otimes\widehat{\Fb}_1(\ub,\vb)\wb\big)\ab=\big(\Fb_1^T(\ub,\vb)\ub\otimes\wb\big)\ab, \\ &\big(\ab\cdot\widehat{\Fb}_2(\ub,\vb)\wb\big)\vb=\big(\vb\otimes\widehat{\Fb}_2(\ub,\vb)\wb\big)\ab=\big(\widehat{\Fb}^T_2(\ub,\vb)\vb\otimes\wb\big)\ab,
\end{aligned}
\end{equation}
we have that
\begin{equation}
(\ab\cdot\wb)\widehat{\fb}(\ub,\vb)-\big( \widehat{\fb}(\ub,\vb)\cdot\wb\big)\ab=\big( \widehat{\Fb}_1^T(\ub,\vb)\ub\otimes\wb+\widehat{\Fb}_2^T(\ub,\vb)\vb\otimes\wb\big)\ab, \quad \quad\forall\ub,\vb,\ab\in\Vc. 
\end{equation}
Given the arbitrariness of $\ab$, it suffices to choose $\ab\cdot\wb=0$ to deduce that
\begin{equation}
\widehat{\fb}(\ub,\vb)\cdot\wb=0, \quad \forall\ub,\vb\in\Vc \quad \Leftrightarrow \quad \widehat{\fb}(\ub,\vb)=f_1(\ub,\vb)\ub+f_2(\ub,\vb)\vb.
\end{equation}
To conclude, one observes that this provisional form of $\widehat{\fb}(\ub,\vb)$ is compatible with \eqref{iso} iff $f_1$ and $f_2$ are isotropic. Thus, an application of \eqref{isosca} suffices.\qed

\section*{Appendix B. Two algebraic lemmas}\label{alglem}
\paragraph{lemma}\label{lemma2}
Let the algebraic equality 
\begin{equation}\label{algeq}
\alpha\Ib+\beta\ub\otimes\ub+\gamma\vb\otimes\vb+\delta(\ub\otimes\vb+\vb\otimes\ub)=\mathbf{0}
\end{equation}
hold for all $\ub,\vb \in{\mathcal V}$. Then, 
\begin{equation}
\alpha=\beta=\gamma=\delta=0.
\end{equation}

\noindent \textit{Proof} Let $\ub,\vb$ be chosen orthogonal and unimodular. Right-multiplication of \eqref{algeq} by  $\ub$ first and then by $\vb$ yields:
\[
\begin{aligned}
&\alpha\ub+\beta\ub+\delta\vb=\mathbf{0},\\
&\alpha\vb+\gamma\vb+\delta\ub=\mathbf{0},
\end{aligned}
\]
whence, given that $\ub\cdot\vb=0$,
\begin{equation}\label{app1}
\alpha+\beta=0, \qquad \alpha+\gamma=0, \qquad \delta=0.
\end{equation}
On the other hand, taking the trace of \eqref{algeq} gives:
\begin{equation}\label{app2}
3\alpha+\beta+\gamma=0.
\end{equation}
To conclude the proof, it is enough to solve  system $\eqref{app1}_{1,2}$-\eqref{app2}.\qed

\paragraph{lemma}\label{lemma4}
Let the algebraic equality 
\begin{equation}\label{algeqt}
\begin{aligned}
&\alpha\Ib+\beta\ub\otimes\ub+\gamma\vb\otimes\vb+\delta(\ub\otimes\vb+\vb\otimes\ub)+\\
&\varphi\eb\otimes\eb+\chi(\ub\otimes\eb+\eb\otimes\ub)+\psi(\vb\otimes\eb+\eb\otimes\vb)=\mathbf{0}
\end{aligned}
\end{equation}
hold for all $\ub,\vb\in {\mathcal V}$. Then, 
\begin{equation}
\alpha=\beta=\gamma=\delta=\varphi=\chi=\psi=0.
\end{equation}

\noindent \textit{Proof} On exploiting the arbitrariness of $\ub$ and $\vb$, we choose $\ub\cdot\vb=0$, $\ub\times\vb=\eb$, $|\ub|=|\vb|=1$. Right-multiplication of \eqref{algeqt} by  $\ub$, $\vb$, and $\eb$, yields:
\begin{equation}
\begin{aligned}
&\alpha\ub+\beta\ub+\delta\vb+\chi\eb=\mathbf{0},\\
&\alpha\vb+\gamma\vb+\delta\ub+\psi\eb=\mathbf{0},\\
&\alpha\eb+\varphi\eb+\chi\ub+\psi\vb=\mathbf{0},
\end{aligned}
\end{equation}
whence
\begin{equation}\label{app3}
\alpha+\beta=0, \quad \alpha+\gamma=0, \quad \alpha+\varphi=0, \quad \delta=\chi=\psi=0.
\end{equation}
Next,  one first writes \eqref{algeqt} for $\ub=\vb$, $\ub\cdot\eb=0$, then applies the resulting tensor to $\ub$, so as to arrive at
\begin{equation}
(\alpha+\beta+\gamma)\,\ub=\mathbf{0},
\end{equation}
with the use of $\eqref{app3}_4$. In view of the arbitrariness of $\ub$, the last relation is tantamount to
\begin{equation}
\alpha+\beta+\gamma=0;
\end{equation}
this, together with $\eqref{app3}_{1-3}$, implies the desired conclusion.\qed

\vskip 12pt

{\scriptsize
\noindent \textbf{Acknowledgements}

\noindent This research was done while AF and PPG visited TU Hamburg-Harburg and Helmholtz-Zentrum Geesthacht. The financial support of the German Science Foundation is gratefully acknowledged.
}


\begin{thebibliography}{00}

\bibitem{BS} S. Bargmann, P. Steinmann, Classical results for a non-classical theory: remarks on thermodynamics relations in Green--Naghdi thermo-hyperelasticity, Continuum Mech. Thermodyn., 19, 1-2, 59-66 (2007).


\bibitem{CN} B.D. Coleman, W. Noll, The thermodynamics of elastic materials with heat
conduction and viscosity, Arch. Rational Mech. Anal. 13, 167-178 (1963).

\bibitem{GN} A.E. Green, P.M. Naghdi, Thermoelasticity without energy dissipation, J. Elast., 31, 189-208 (1993).

\bibitem{GNRoy} A.E. Green, P.M. Naghdi,  A re-examination of the basic postulates of thermomechanics. Proc. R. Soc. Lond. A 432, 171-194 (1991).

\bibitem{Liu} I.-S. Liu, Method of Lagrange multipliers for exploitation of the entropy principle, Arch. Rat.  Mech. Anal., 46, 131-148 (1972).

\bibitem{Liu1} I.-S. Liu, On the entropy supply in a classical and relativistic fluid, Arch. Rat. Mech. Anal., 50, 111-117 (1973).

\bibitem{Liu2} I.-S. Liu, On entropy flux of transversely isotropic elastic bodies, J. Elast. 96, 2, 97-104 (2009).

\bibitem{Mu} I. M\"{u}ller, The coldness, a universal function in thermoelastic bodies, Arch.Rat.Mech.Anal., 41, 319-332 (1971).

\bibitem{ITD} I. M\"{u}ller,\emph{Thermodynamics}. Pitman, 1985.

\bibitem{PPG} P. Podio--Guidugli, Untitled, forthcoming (2012).

\end{thebibliography}
\end{document}